\documentclass[12pt,preprint]{aastex}

\usepackage{graphicx}
\usepackage{xcolor}
\usepackage{verbatim}

\shorttitle{OH Maser Survey II: Satellite Lines}
\shortauthors{Ruiz-Velasco, et~al.}
\begin{document}

\title{VLBA SURVEY OF OH MASERS IN STAR-FORMING REGIONS II: SATELLITE LINES}
\author{A. E. Ruiz-Velasco\altaffilmark{1}, D. Felli\altaffilmark{2}, V. Migenes\altaffilmark{2, 3}  \and B. K. Wiggins\altaffilmark{2,4}
}

\altaffiltext{1}{Lowell Observatory, 1400 West Mars Hill Road, Flagstaff, AZ 86001, USA. e-mail: alma@lowell.edu} 
\altaffiltext{2}{Brigham Young University, Department of Physics and Astronomy, ESC, Provo, Utah 84602}
\altaffiltext{3}{Texas Southern University, Department of Physics, 3100 Clebourne Ave, Houston, TX. 77004}
\altaffiltext{4}{Los Alamos National Laboratory, P.O. Box 1663, Los Alamos, NM 87545}
\begin{abstract}

Using the Very Long Baseline Array (VLBA) we performed a high resolution OH maser survey in Galactic star-forming regions (SFRs). We observed all the ground state spectral lines: the main lines at 1665 and 1667 MHz and the satellite lines at 1612 and 1720 MHz. 
Due to the exceptionality of finding satellite lines in SFRs, we will focus our discussion on those lines.
In our sample of 41 OH maser sources, 
five (12\%) showed the 1612 MHz line and ten (24\%) showed the 1720 MHz line, with only one source showing both lines.
We find that 1720 MHz emission is correlated with the presence of HII regions, suggesting that this emission could be used to diagnose or trace high-mass star formation.
We include an analysis of the possible mechanisms that could be causing this correlation as well as assessing the possible relationships between lines in our sample. In particular, the presence of magnetic fields
seems to play an important role, as we found Zeeman splitting in four of our sources (W75~N, W3(OH), W51 and NGC~7538).
Our results have implications for current understanding of the formation of high-mass stars as well as on the masing processes present in SFRs.

\end{abstract}


\section{Introduction}

Masers can serve as remarkable probes of conditions in star-forming regions (SFRs).
Gas velocities, temperatures, densities and magnetic field strengths can be inferred from observations of maser transitions.  Masers provide a unique window to study the conditions and dynamics of massive-star formation which remains a critical open question in the literature.

Observations of OH masers in SFRs have been performed mainly in the 1665 and 1667 MHz lines, or so-called ``main'' OH emission lines. OH satellite lines (1612 MHz and 1720 MHz) are not normally observed and are considered rare. 
However, satellite lines trace particularly interesting astrophysical processes; the 1612 MHz line is associated with circumstellar envelopes of OH/IR stars and Mira-type stars \citep[e.g.,][]{HermanHabing85, david93} while 1720 MHz emission appears in shocked, compressed gas in supernova remnants \citep[e.g.,][]{Gray2012, frail96}. Here we study the appearance of such lines in the context of SFRs where the theory of OH satellite maser lines has not been developed in detail (see Caswell 1999). Understanding the relationship between the satellite lines and the main OH lines as well as their connection to conditions in their environments will further enable their use as diagnostics of SFRs.

In a survey of 200 OH maser sources in the southern sky, \citet{Caswell2004} reported that only about 
17\% of the sources presented the 1720 MHz line. All sample members in \citet{Caswell2004} exhibited 1665 MHz emission, which would not have allowed the authors to constrain the dependence of the other lines on the appearance of the 1665 MHz line. In this work, we present results for a survey for 1667 MHz, 1665 MHz, 1612 MHz and 1720 MHz emission in 41 Galactic radio sources, many of which are established compact HII regions. Our survey provides an opportunity to assess correlations between satellite and main lines in high-mass, star-forming regions via categorical data analysis. The sample size in our survey is also sufficiently large to revisit the detection rates of satellite lines in SFRs and to provide some comment on the environments in which such emission arises.

This paper is the second in a series of papers that describe the results of a large OH maser survey, and is the continuation of \citet{Migenes2016}.  
Preliminary results regarding the main lines only were published in \citet{Ruizv2006}. This work is structured as follows. In \S\ 2 we detail our observations. We summarize survey results in \S\ 3 and discuss particularly noteworthy sources in \S\ 4. In \S\ 5, we provide a discussion on the dependence of 1720 MHz emission on magnetic field strength as well as a statistical analysis of our results. We describe our conclusions in \S\ 6.

\section{Observations}

We selected 41 Galactic radio sources observable in the northern hemisphere, most
of them previously known to be compact HII regions.
Observations were conducted on January 1, 2001 with the
NRAO\footnote{The National Radio Astronomy Observatory is a facility of the National Science
Foundation operated under cooperative agreement by Associated Universities,
Inc.} Very Long Baseline Array (VLBA), using all 10 antennas.
The data was observed at the 250 kHz bandwidth and processed with 256 channels
so that a spectral resolution of 0.98 KHz (0.17 $\rm{km\,s}^{-1}$) was achieved.
We observed the four ground state OH maser transitions (1665, 1667, 1612 and 1720 MHz) in
right and left circular polarization modes.

The VLBA provided us with an angular resolution of 4.3 mas. We used the VLBA in a snap-shot mode, observing every source with a six-minute scan. Using the VLBA in this manner provided a sensitivity of $\sim$ 60 mJy beam$^{-1}$, and an rms noise of $\sim$ 0.5 Jy.
The data was calibrated and processed using the NRAO \texttt{AIPS} software package
with the spectral line VLBA data reduction standard protocol.
We detected fringes for approximately $70\,\%$ of the sources.
The continuum sources NRAO~512, BL~LAC, CTA~102, J2007+4029 and J2005+7752
were all observed at 1.6 GHz for delay calibration and bandpass correction.

%%%%  RESULTS  %%%%%
\section{Results}

We detected OH maser emission for at least one line in 32 of 41 sources. Fifteen of these 32 sources possessed highly compact structures and strong ($ > 1$ Jy) emission: IRAS\,18032$-$2032, W31, IRAS\,18117$-$1753,  IRAS\,18507$+$0110, W48, IRAS\,03558$-$0003, W51, W75~N, W75~S,  IRAS\,21413$+$5442,  IRAS\,22176$+$6303, Cepheus~A, NGC~7538, W3(OH) and IRAS\,06117$+$1350.  Nine of the sources (IRAS\,01190$-$0014, IRAS\,01961$-$0023, \linebreak IRAS\,03124$-$0011, IRAS\,03140$-$0026, IRAS\,03274$-$0008, IRAS\,03313$-$0009,  AFGL~2591, IRAS\,21306$+$5539, IRAS\,00338$+$6312) were not detected at any frequency. It is possible that these sources were completely resolved or that the integration time was too short for detection.

We found satellite lines in 14 of the 41 sources in this survey (34\%), with five (12\%) showing the 1612 MHz line and ten (24\%) showing the 1720 MHz line. Only one source, W3(OH), showed both lines. Essentially, all the sources displaying the 1720 MHz line also have the 1665 MHz line. The only exception, IRAS\,03129$+$0007, was marginally detected ($S_{\mbox{\tiny 1720 MHz}} = 0.3$ Jy) and hence may be an outlier. The sources IRAS\,01222$-$0012, GRS\,2086$+$048 and IRAS\,03060$-$0004 show the 1612 MHz line only. W3(OH) is the only source showing all the OH maser transitions (see \S~\ref{sec:sources}).

In Table~\ref{tab:detections} we have summarized the results of the survey. %in ascending order according to the Galactic coordinates. 
Column (1) includes the common name or the IRAS name; column (2) the Galactic coordinates; 
columns (3) and (4) the equatorial coordinates (equinox J2000); column (5) the systematic velocity with respect to the local standard of rest in $\rm{km\,s}^{-1}$; and columns (6) - (9) the flux in janskys of the highest peak for each frequency observed. We also indicate the polarization in parentheses, where R and L are right and left circular polarization, respectively.

\section{Interesting Sources}
\label{sec:sources}

The following are widely known high-mass SFRs showing either one or both satellite lines. Additionally, they present a spectrum very similar to that of previous observations in the main lines, indicating that the emission is compact enough to be observed with higher resolution instruments.  

\subsection{W75 North} 
\label{sec:w75n}

W75 North (hereafter W75~N) is a well-known region of active high-mass star formation located at a distance $\leq$ 2 kpc \citep{odenwald93}. It contains three main radio continuum sources in different stages of evolution. One of them, VLA~2, harbors
an ultra-compact HII region \citep{Torrelles97} and has associated masers of different species: water vapor \citep{LekhtKrasnov2000, Torrelles2003}, 
hydroxyl \citep{Haschick1981} and methanol \citep{LiechtiWilson96}.

W75~N harbors a large-scale, bipolar outflow \citep{Surcis11}
and displayed an OH maser flare of more than 1000~Jy at 1665 MHz in 2003 \citep{alakoz05}. 
Its radial velocity of 2 $\rm{km\, s}^{-1}$ and position were strongly suggestive of an association with the source VLA~2.
This OH maser flare was the first such flare ever reported, with an increase in the intensity of two orders of magnitude in a period of  less than six  years.
\citet{Slysh2010} proposed that the flare was produced by a magnetohydrodynamic shock propagating along a filament traced by more OH maser spots.

\begin{figure}
\includegraphics[angle=270,scale=0.50]{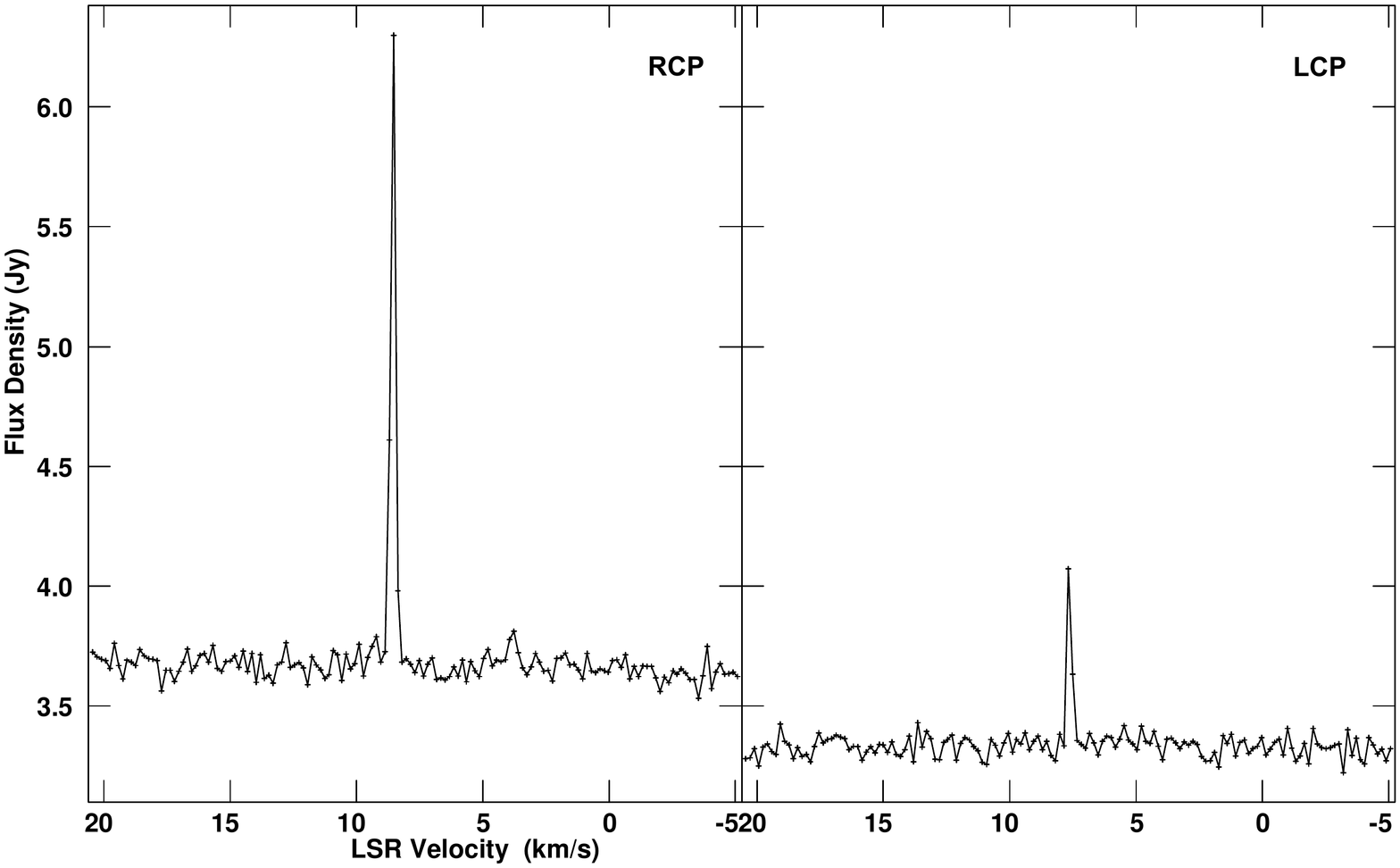}
\caption{W75~N at 1720 MHz. The horizontal axis is the local standard of rest (LSR) velocity. The vertical axis is the flux density in janskys. From left to right, the first panel shows the right circular polarization (RCP) and the second panels shows the left circular polarization (LCP).}
\label{fig:w75n}
\end{figure}

Our 1665 MHz spectrum shows five strong features spread over a velocity range of 15 $\rm{km\,s}^{-1}$, where the strongest feature has a flux of  62~Jy in RCP and 130~Jy in LCP both at a radial velocity of 2.4 $\rm{km\,s}^{-1}$.  
At 1667 MHz we identify seven features, the strongest at 9.3 $\rm{km\,s}^{-1}$ with 11 Jy (RCP) and 16 Jy (LCP).
At 1720 MHz we have a single peak with a velocity of 7 $\rm{km\,s}^{-1}$ and strength of 6.4 Jy (RCP) and 4.1 Jy (LCP) above the noise (see Fig.~\ref{fig:w75n}). Since our observations date from 2001, the flux we measure is considered to be the precursor of the 2003 flare.

\subsection{W51} 
\label{sec:w51}

Located at $\sim5.4$ kpc, the W51 complex hosts some of the strongest H$_2$O maser sources in the Galaxy \citep{sato2010}. 
In W51 Main, two powerful bipolar outflows have been detected; the high-velocity one (W51e2-E) coincides with the CO $(J=2\relbar1)$ emission observed by \citet{KetoKlaassen08}, which is generated by a young and massive star of $M_\star > 15 M_\odot$. The low-velocity outflow (W51e2NW) is almost perpendicular to the former; it is not clear if they are associated. In addition, the supernova remnant W51C lies $22.75\arcmin$ south \citep{brogan2013}, though any interaction with W51 Main seems unlikely.

\begin{figure}
\includegraphics[angle=270,scale=0.5]{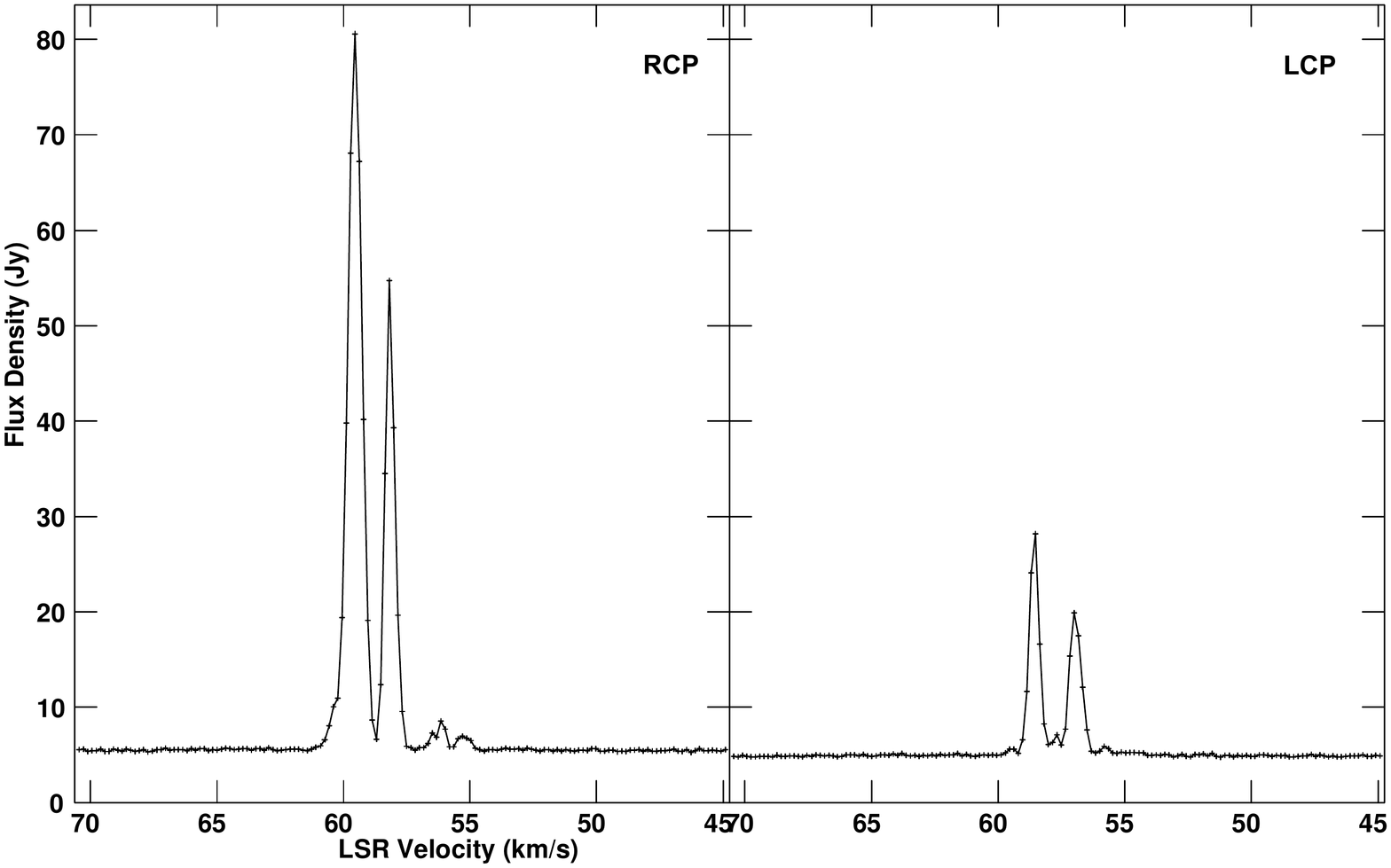}
\caption{W51 at 1720 MHz. The horizontal axis is the local standard of rest (LSR) velocity. The vertical axis is the flux density in janskys. From left to right, the first panel shows the right circular polarization (RCP) and the second panels shows the left circular polarization (LCP).}
\label{fig:w51}
\end{figure}

We detected one strong spectral feature in the 1665 MHz line with a width of 0.7 $\rm{km\,s}^{-1}$. Considering that we are observing an OH maser line with a high-resolution radio interferometer, this feature seems particularly broad. The feature has a flux of 67 Jy in RCP and 24 Jy in LCP, and a radial velocity of 58 $\rm{km\,s}^{-1}$ in both polarizations. At 1667 MHz we see one feature with a peak of 0.6~Jy and a radial velocity of 62 $\rm{km\,s}^{-1}$ in RCP and two features in LCP: a 3~Jy peak at 54 $\rm{km\,s}^{-1}$ and another of $\lesssim$ 1~Jy at 62 $\rm{km\,s}^{-1}$.
The 1720 MHz line shows a strong double peak between 55 and 60  $\rm{km\,s}^{-1}$ with corresponding fluxes of 81.2 and 54.5~Jy (RCP) 
and 28.5 and 20.2~Jy (LCP) above the rms noise (see Fig.~\ref{fig:w51}). The double peak shifts slightly when we change the polarization, indicating a Zeeman splitting (see Sec~\ref{sec:magfields}). 

Our observations on the W51 Main complex covered both the high-velocity (W51e2-E) and the low-velocity (W51e2-NW) outflows. We presume at least one of these outflows produces the collisional pumping required to see the 1720 MHz line.

\subsection{NGC\,7538} 
\label{sec:ngc7538}

Located at a distance of $\sim 2.6$ kpc \citep{moscadelli09} this massive star-forming region harbors a hypercompact HII region \citep{sewilo04}. 
According to \citet{Beuther12}, at least 40 M$_\odot$ (and potentially even more than 100 M$_\odot$)
are concentrated within approximately 2000 AU.

Using 6.7 and 12.2 GHz methanol maser observations, \citet{Minier2000} concluded that the masers trace what seems to be an edge-on disk of length $\sim$280 AU. With sub-millimeter maps of the CO emission \citet{qiu2011} concluded that the elongated structure seen by \citet{Minier2000} was instead better described by multiple outflows mainly coming from a young $\sim$ 20 M$_\odot$ O star. It is interesting to note that our observations coincide in space as well as in the velocity field with this O star and add further insight into Minier's and Qiu's work.

\begin{figure}
\includegraphics[angle=270,scale=0.5]{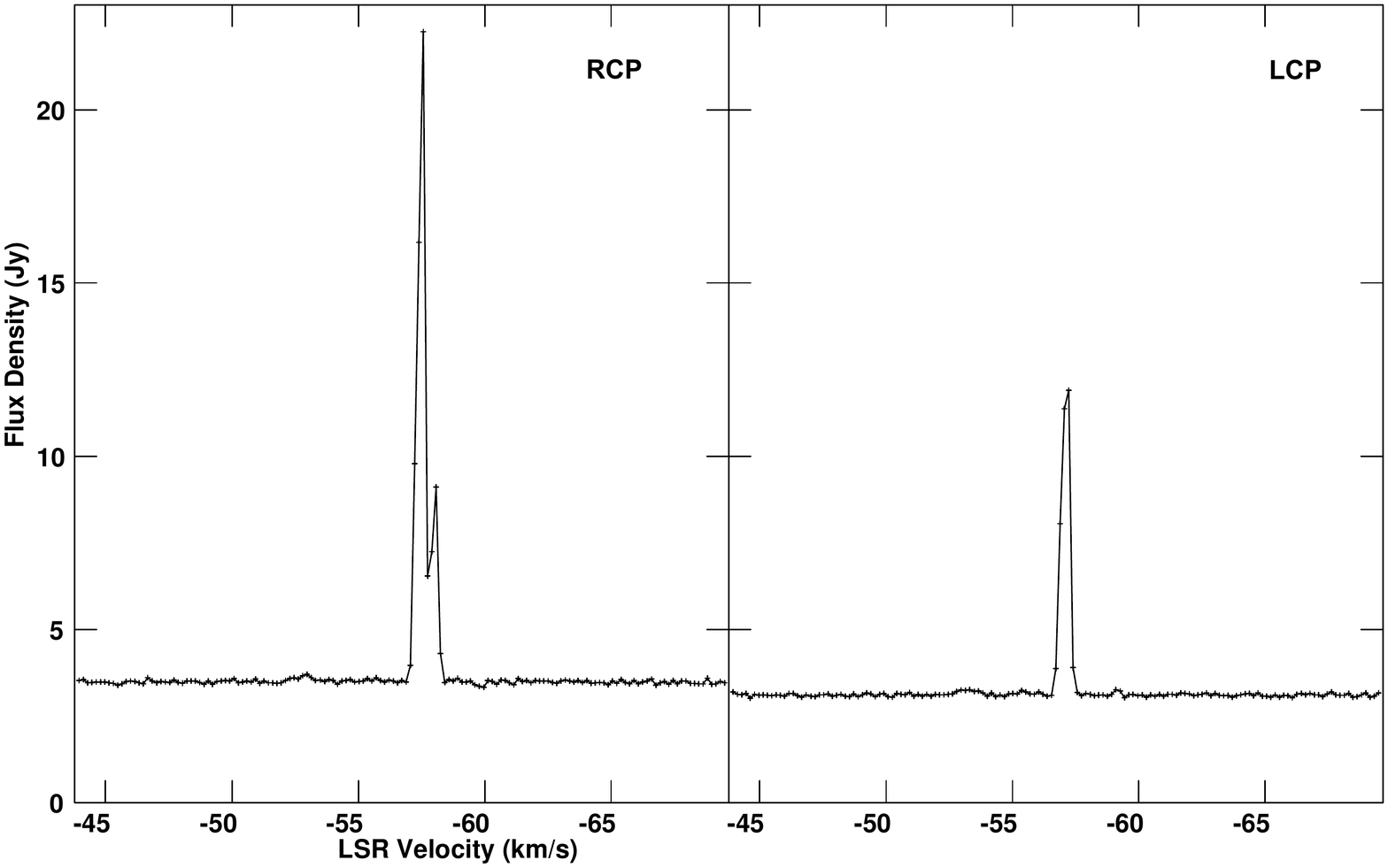}
\caption{NGC~7538 at 1720 MHz. The horizontal axis is the local standard of rest (LSR) velocity. The vertical axis is the flux density in janskys. From left to right, the first panel shows the right circular polarization (RCP) and the second panels shows the left circular polarization (LCP).}
\label{fig:ngc7538}
\end{figure}

We see a single feature in 1665 MHz at the radial velocity of $-$59.5 $\rm{km\,s}^{-1}$, with a flux density of 0.5 Jy in RCP and 6 Jy in LCP. 
At 1667 MHz (RCP) there is one feature with a flux density of 1.4 Jy at the radial velocity of $-$64.4 $\rm{km\,s}^{-1}$. 
At 1720 MHz we found a broad feature with a radial velocity of $-$57.4 $\rm{km\,s}^{-1}$, which appears to be double-peaked when observed in the RCP spectrum. It has a flux density of 22.6 Jy (RCP) and 11.9~Jy (LCP) above the rms noise (see Fig.~\ref{fig:ngc7538}).

\subsection{W3(OH)}
\label{sec:w30h}

W3(OH) is a surprisingly hard X-ray source located at a distance of $\sim1.9$ kpc \citep{choi2014}, and it is the only emitter in our sample to exhibit both the 1612 and the 1720 MHz transitions. 

\citet{Fish2006} found more than 250 ground-state OH masers and 56 Zeeman pairs in this source.
Using the Chandra X-ray Observatory, \citet{FeigsonTownsley2008} found a heavily obscured, 
young massive star ionizing a region where the masers are observed.
At 1665 MHz the spectrum shows six features, with the strongest at $-$45 $\rm{km\,s}^{-1}$ with a flux of 140 Jy (RCP). In 1667 MHz there is a broad line with a flux density of 14~Jy in LCP.

\begin{figure}
\includegraphics[angle=270,scale=0.5]{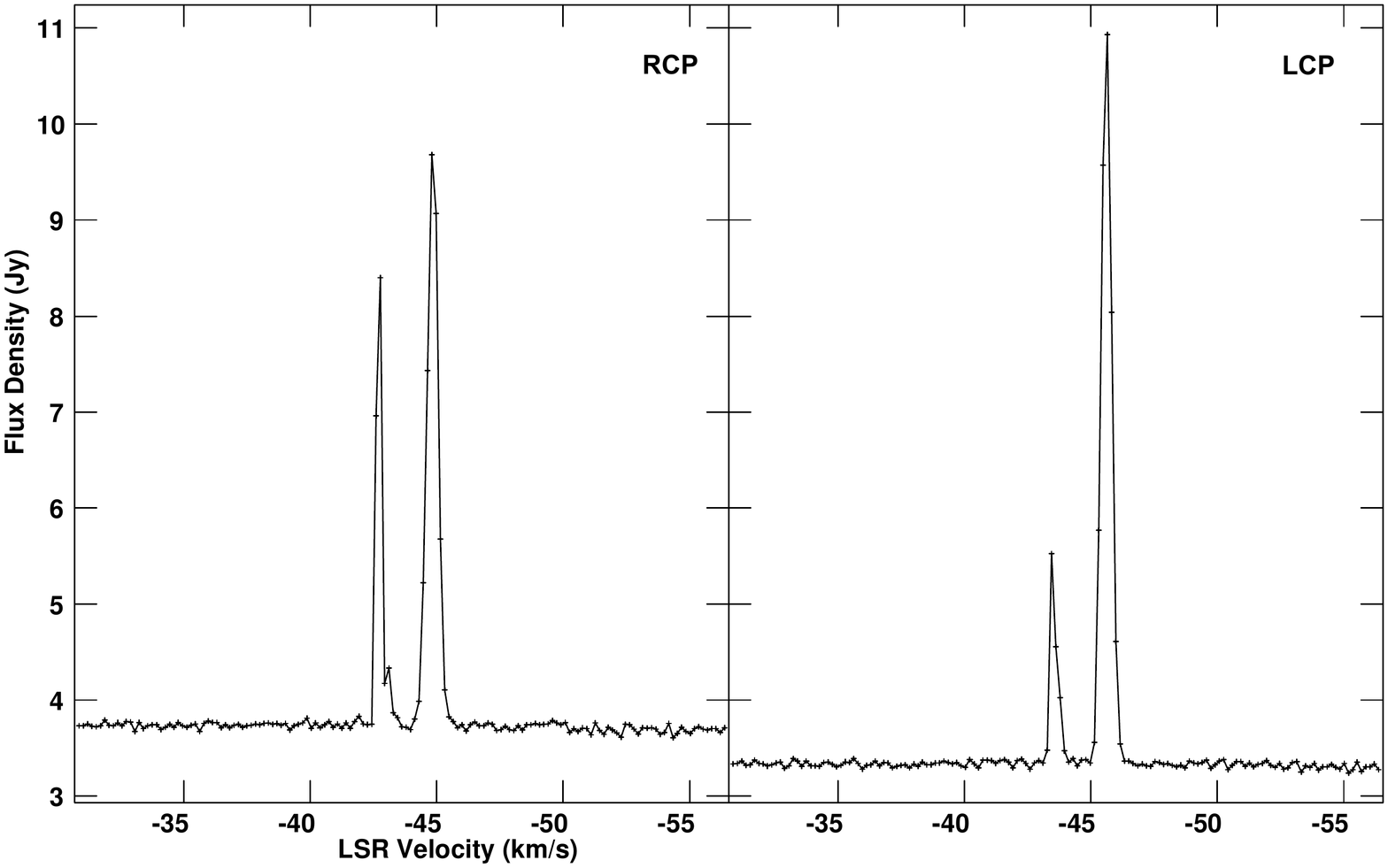}
\caption{W3(OH) at 1720 MHz. The horizontal axis is the local standard of rest (LSR) velocity. The vertical axis is the flux density in janskys. From left to right, the first panel shows the right circular polarization (RCP) and the second panels shows the left circular polarization (LCP).}
\label{fig:w3oh1720}
\end{figure}

\begin{figure}
\includegraphics[angle=270,scale=0.5]{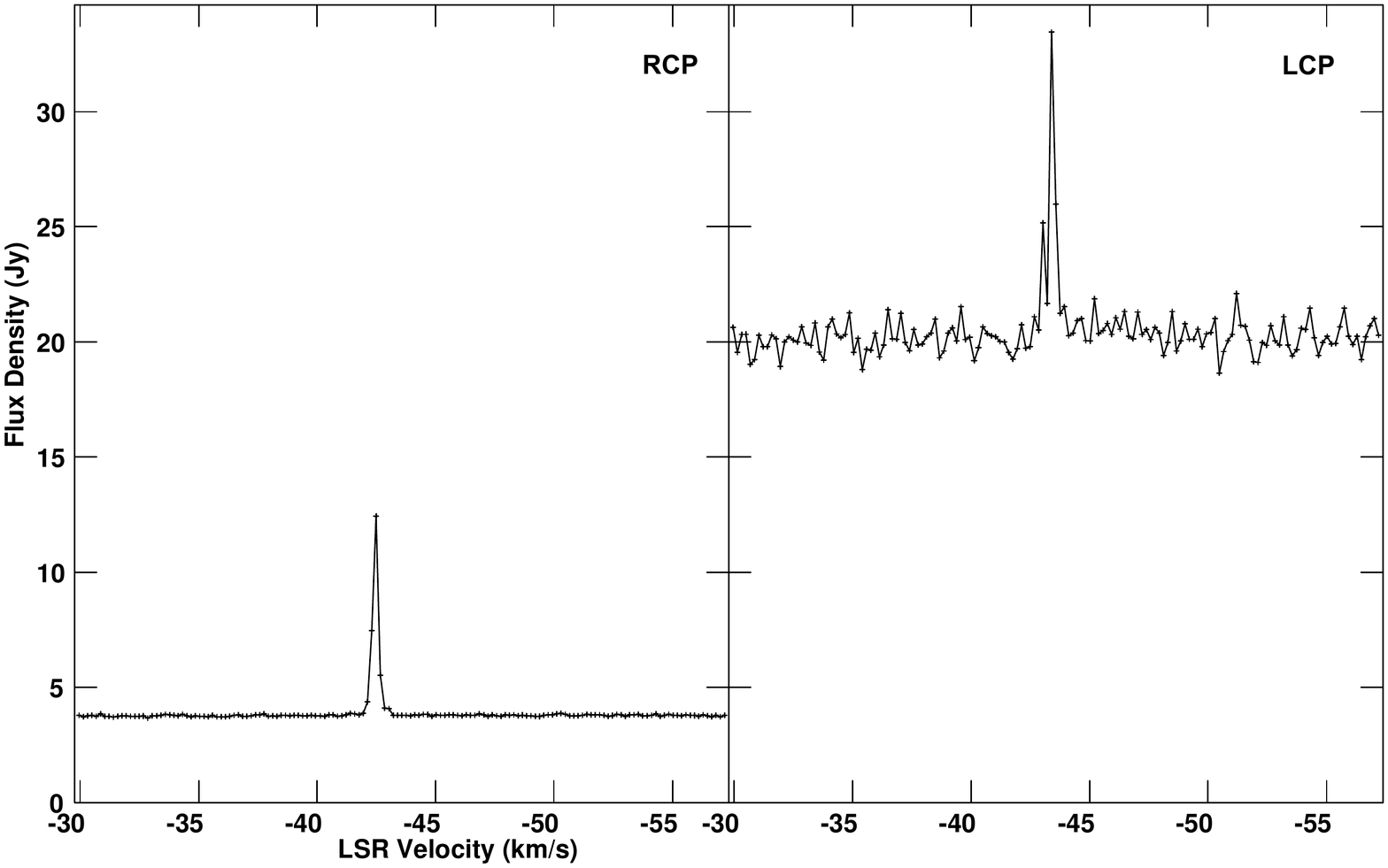}
\caption{W3(OH) at 1612 MHz. The horizontal axis is the local standard of rest (LSR) velocity. The vertical axis is the flux density in janskys. From left to right, the first panel shows the right circular polarization (RCP) and the second panels shows the left circular polarization (LCP).}
\label{fig:w3oh1612}
\end{figure}

The 1612 MHz line shows one feature at 22.8~Jy (RCP) and 51.1~Jy (LCP) above the rms noise (see Fig.~\ref{fig:w3oh1612}).  
The 1720 MHz line has a double feature with corresponding fluxes of 12.8 Jy and 15.4~Jy in RCP, and 7.7~Jy and 19.2~Jy in LCP respectively (see Fig.~\ref{fig:w3oh1720}). 
Even though both features have a radial velocity of about $-$44 $\rm{km\,s}^{-1}$, \citet{Fouquet} affirm they arise from different spots.

%%  Discussion  %% 
\section{Discussion}
\label{sec:discussion}

The  conditions for the satellite lines to occur are very specific:
in the case of the 1612 MHz line, the population inversion results mainly from radiative pumping with far-infrared photons and temperatures around 50 K \citep{Elitzur92}.
The 1720 MHz line can be achieved by either collisional or radiative pumping. If the temperature is between 25 and 200 K and the OH density is $n_{\mbox{\tiny OH}} \geq 10^{-3}-10^{-4}\mbox{ cm}^{-3}$,
collisions (with H$_2$ molecules) produce a strong population inversion of the 1720 MHz line \citep{Elitzur76}.
This becomes particularly meaningful for our sample since the outflows observed in some of the sources may be triggering the collisions. For example, in NGC~7538
\citet{Beuther12} suggest that an outflow 
must be very close to the line of sight, allowing the infrared radiation to escape through the outflow cavity. 
This scenario could explain why the 1720 MHz 
%line seems to come directly from the compact source instead of from the outflows projected northeast of it. 
masers are coincident with the compact source, instead of being observed closer to the multiple outflows detected northeast of it.

On the other hand, with lower OH densities and a strong infrared field, both lines can be produced, with the 1612 MHz line dominating. This seems to be the case of W3(OH)  where the satellite OH masers are located slightly offset from an infrared source of Class 0/I at (J2000) R.A.$=02^h 27^m 03\fs86$ Dec$=61\arcdeg 52\arcmin 25\farcs32$ \citep{Rivera-Ingraham2011}.

\subsection{Magnetic Fields and the 1720 MHz Line}
\label{sec:magfields}

We noticed that all of the sources displaying the 1720 MHz line appear to have at least some degree of polarization. 
This is of great significance since in the presence of a magnetic field the molecules will move along the field lines. Additionally, the outflow  will produce collisions and enhance the density of the medium.

In particular, W75~N presents an interesting association. As mentioned above (in \S~\ref{sec:w75n}), W75~N exhibited a huge flare in the 1665 MHz line in 2003. The pre-flare observations taken in 2001 reported a strong magnetic fields of 40 mG \citep{SlyshMigenes2006}, increasing to 70 mG at the time of the flare \citep{Slysh2010}. 
In this same region \citet{Surcis11} measured the magnetic fields using H$_2$O masers, finding values of $\sim$ 700 mG in VLA~1 and $\sim$ 1700 mG in VLA~2. They conclude this is an indication of shock compression of the gas. More recently, \citet{charly2015} published long-term VLA observations of this same source. They found that the magnetic field around VLA~2 changed its orientation, following the direction of the large-scale magnetic field in the region. Moreover, they observe the transition from an uncollimated outflow to a collimated one during the 18 years of observations.

Careful analysis of the 1720 MHz data on W75~N indeed indicates the presence of a magnetic field of 1.7~mG. Additionally, we look for Zeeman pairs in the rest of the sources mentioned above and found values of 2.1~mG for W51 and 1.4~mG for W3(OH).
In the source NGC~7538, the Zeeman pair is less clear, given that the line seems to be an unresolved double peak. 
We estimate a value of 0.7~mG for this source.

\citet{Fish2003} performed an OH-maser Zeeman pair survey on massive SFRs in the Galaxy, including the satellite lines for some sources, where W51 and W3(OH) were among them. 
For W51 they found 5.0, 4.0, 3.0 and 3.5 mG at 1720 MHz and for W3(OH) they found 5.3 mG at 1720 MHz. Our results are consistent with these values.

\citet{Fish2003} also suggest that 
the physical requirements for masing at 1612 and 1720 MHz cause satellite-line maser spots to be observed preferentially at locations of increased magnetic field strength. 

There are strong reasons to consider magnetic fields in the high-mass star formation process.
Using radiation-magnetohydrodynamical simulations, \citet{Peters2011} produced 50\% more massive stars compared to the equivalent simulation without magnetic fields. They found that
stars can grow larger masses mainly because during the formation phase, magnetic fields work against gravitational collapse. This delays fragmentation and also increases the accretion rate as they drain the angular momentum in the center, feeding more material to the protostar.

All the conditions discussed above (high-density medium, collisions, outflows and magnetic fields) favor the pumping of satellite lines, and hence the presence of these lines could be used as an indicator of ongoing high-mass star formation.

\subsection{Statistical Analysis}
\label{sec:stats}

\cite{Caswell2004} detected 1720 MHz emission among 17\% of 200 established 1665 MHz
OH SFRS. The number of 1720 MHz sources detected in our survey, given 1665 MHz 
emission, is 9 or $\sim 30$\%, an increase from \cite{Caswell2004}.  To test if our figure
differs significantly from \cite{Caswell2004}  we first carried out a simple non-parametric 
$\chi^2$ test for equality of proportions. We obtained a $p$-value of $p = 0.07823$ 
indicating that differences in our detection of the 1720 MHz line among 1665 MHz OH
maser SFRs is marginally statistically significant. Under the assumptions 
of a one-tailed parametric proportions test, i.e. that numbers of detections of the 1720 MHz line
among 1665 MHz maser SFRs may be safely considered approximately normally distributed, the probability of 
obtaining this higher proportion of 1720 MHz detections by chance is
$p = 0.03911$, which is sufficient to reject the hypothesis that our detection
rate is less than or equal to that of \cite{Caswell2004} at 5\% significance. Our survey may demonstrate, therefore,
that detections of the 1720 MHz satellite line in regions shared by
1665 MHz masers could be larger than the 17\% figure in \cite{Caswell2004}.

\cite{Cas99} reported nearly equal numbers of detections for both satellite lines in SFRs with 1665 MHz emission. This trend is not reflected in our data. If we consider only sites with 1665 MHz emission, we assess the probability of achieving our respective proportions of detections and non-detections for equally probable 1720 MHz and 1612 MHz lines to be $p = 0.043$. 

\cite{Caswell2004} examined a sample of galaxies all exhibiting at least 1665 MHz emission. Our data provided an opportunity to statistically explore relationships between lines as we do not have a similar restraint on members of our sample. Reducing the last four columns of Table~\ref{tab:detections} to binary variables (detections vs. non-detections) allows for a categorical analysis.  We first assemble a $2 \times 4$ contingency table to assess if detection rates differ significantly between lines (Table~\ref{tab:contingency}).
%detection rates between the lines (5:3:1:2) for 1665, 1667 1612 and 1720 respectively

We applied a $\chi^2$ goodness-of-fit test using Yates continuity correction \citep{Yates34} for small sample sizes to evaluate the null hypothesis that detection rates are identical between all lines. This yielded $\chi^2 = 28.703$, corresponding to a very small $p$-value of $p=2.585\times10^{-6}$, which is very strong evidence for the established belief that detection rates between lines are not identical.

We also carried out a logistic regression \citep[see][]{Ag02,Christensen90} to explore potential correlations between lines. We set up 12 Generalized Linear Models (GLMs), one for each pair of lines, each having the form: 

\begin{equation}
\log \left( \frac{p_i}{1- p_i} \right) = \beta_0 +  \beta_{ij} X_j,
\end{equation}
where $i$ and $j$ range over the lines, i.e. 1665 MHz, 1667 MHz, 1612 MHz and 1720 MHz, and $p$ is the empirical probability of the $i$th line calculated from Table \ref{tab:detections}.  The coefficients $\beta_i$ are determined through a maximum likelihood fitting routine in R (R Core Team 2015)\footnote{R Core Team, 2013, R: A language and environment for statistical computing. R Foundation for Statistical Computing, Vienna, Austria.}. The dichotomous predictive variables $X_j$ are assumed to be independent. Our coefficient fits for each of our models within a 95\% confidence interval appear in Table \ref{tab:pair_log}.  To explore the dependency of each line of the combination on all other lines, we also set up 4 GLMs, one for each line, each of the form: 

\begin{equation}
\log \left( \frac{p_i}{1- p_i} \right) = \beta_0 + \sum_{i \neq j} \beta_j X_j,
\end{equation}
i.e. the sum is taken over the remaining three lines. These results appear in Table \ref{tab:Log}.

With few exceptions we note our data is insufficient to meaningfully constrain coefficients. The probability that observed trends, however noisy, arise from random variations in fundamentally uncorrelated data is also reported. The $p$-values should be interpreted as the significance that variation in the independent variables $X_j$ predicts or explains variation in the response variable $\log \left( \frac{p_i}{1- p_i} \right)$. In the pairwise logistic fits, several pairs show statistically significant correlations including the two main OH lines at 1665 and 1667 MHz and the satellite line at 1720 MHz with each of the main lines. We find no correlations involving the 1612 MHz line. In contrast, our GLM, where the presence of a maser is a function of the presence of all other lines yields only one statistically significant correlation between the main 1665 and 1667 MHz lines.  In principle, if the upper $\Lambda$ doublet of the ground state $\Pi_{3/2} (J = 3/2)$ is inverted with respect to the lower doublet, we might expect all four transitions to be masing simultaneously, but this is not reflected in our data, nor is it observed generally 
\citep[see][]{lo05}, indicative of a complex pumping mechanism. Our data robustly identifies the well-known relationship between the two major OH maser lines with additional statistical evidence from the pairwise analysis for correlations of the 1720 MHz line with the 1665 MHz and 1667 MHz lines.

If we use object classifications as given in SIMBAD and \cite{Migenes2005}, it appears that 1720 MHz emission is correlated with objects classified as HII regions. If we carry out a logistic fit to our 1720 MHz detections and objects of HII type, the 95\% confidence interval for our coefficient for 1720 MHz emission is [-0.157, 2.78] with a 
$p$-value of 0.0891. This corresponds to a marginally significant ($\alpha < 0.1$), positive correlation and suggests that for each additional 1720 MHz detection we expect on average $\sim 1$ additional HII region to appear. Our $p$-value provides the interesting implication that 1720 MHz emission predicts with non-negligible levels of confidence the presence of HII regions in our data. No correlation of significance is found with 1612 MHz line ($p = 0.379$), possibly due to the fact that our sample size contains a very small number (five) of 1612 MHz detections. Additional surveys with longer integration times will be needed to successfully constrain any connection between the 1612 MHz emission and star formation processes.

%%% Conclusions  %%%
\section{Conclusions}

We present a high-resolution OH maser survey with the VLBA toward 41 Galactic SFRs.
Out of the 41 OH maser sources observed, five (12\%) showed the 1612 MHz line and ten (24\%) showed the 1720 MHz line.

Compared to previous surveys \citep{Caswell2004,sevenster97} we detect a higher proportion of satellite lines in a smaller sample, which we find to be statistically significant in parametric tests. Though we cannot rule out the effects of selection bias in our survey, our data could indicate that satellite lines in SFRs may be more common than previously believed. We additionally find that 1720 MHz emission is correlated with 1665 MHz and 1667 MHz emission. The 1720 MHz line correlates non-negligibly with the presence of HII regions which could suggest that such emission could be used to trace high-mass star formation activity.  We find no statistical evidence for such a correlation with the 1612 MHz satellite line. This is possibly due to the small number of detections (five) in our survey.  

% Updated the statistical analysis to probe this point.
%The 1720 MHz line correlates well with high-mass star forming regions whereas the 1612 MHz line showed a weaker correlation but still only appeared in sources that are previously known to harbor HII regions.
%
We understand that high-mass SFRs contain different physical conditions which allow for strong maser emission of the OH satellite lines,
and in particular, strong magnetic fields may be triggering the 1720 MHz line. 
The future study of these sources at higher resolution and (or) longer integration time will provide more evidence for our results.
    
\acknowledgments

A. R. V. acknowledges the support of CONACYT and Lowell Observatory. D. F. would like to acknowledge a BYU ORCA grant in helping fund the research project.
This research has made use of the SIMBAD database, operated at CDS, Strasbourg, France, and NASA's Astrophysics Data System.
{\it Facilities:} \facility{VLBA (NRAO)}.

\clearpage

%%%%%%%%%%  All Detections Table %%%%%%%%%%%%%%%%%%%%%%%
\begin{deluxetable}{lrcrccccc}
\tablecolumns{11}
\tabletypesize{\scriptsize}
%\rotate
\tablecaption{OH Maser Survey Detections\label{tab:detections}}
\tablewidth{0pt}
\tablehead{ \colhead{(1)} & \colhead{(2)} & \colhead{(3)} & \colhead{(4)} & \colhead{(5)} & \colhead{(6)} & \colhead{(7)} & \colhead{(8)} & \colhead{(9)}  \\
\colhead{Name} & \colhead{Galactic}  & \colhead{R.A.} & \colhead{Dec.} & \colhead{V$_{\mbox{\tiny LSR}}$} & \multicolumn{4}{c}{Flux Density (Jy) and Polarization\tablenotemark{\dag}} \\
& \colhead{Coordinates}  &\colhead{(J2000)} & \colhead{(J2000)} & 
\colhead{(km s$^{-1}$)} & \colhead{1665 MHz} & \colhead{1667 MHz} & \colhead{1612 MHz} & \colhead{1720 MHz}  
}
\startdata
IRAS\,17574$-$2403 & 5.88$-$0.39  & 18~00~34.4 & $-$24~04~04 &14.0 & 0.6(L) & 0.5(R, L) &  x & x \\
CRL\,2059 & 6.05$-$1.45  & 18~04~53.9 & $-$24~26~41 & 11.0 & 2.0(R, L) & x &  x & x \\
IRAS\,18032$-$2032 & 9.62$+$0.20  & 18~06~14.7 & $-$20~31~31 & 1.8 & 1.5(R, L) & 1.4(L) &  x & x  \\
W31 & 10.62$-$0.38  & 18~10~28.6 & $-$19~55~51 & $-$2.2 & 5.0(R, L) &  9.0(R, L) &  x & x  \\
IRAS\,01190$-$0014 & 11.91$-$0.15  &18~12~11.3 & $-$18~41~30 & 40.5 & x & x &  x & x  \\
IRAS\,01222$-$0012 & 12.22$-$0.12  & 18~12~44.5 & $-$18~24~25 & 27.0  & x & x &  0.25(L) & x  \\
W33B  & 12.68$-$0.18 & 18~13~54.7 & $-$18~01~46 & 61.5 & 0.6(L) & x &  x & 0.5(R)  \\
IRAS\,01289$+$0049 & 12.89+0.49  & 18~11~51.3 & $-$17~31~29 & 35.0 & 1.8(R, L) & x &  x & x  \\
IRAS\,18117$-$1753 & 12.91$-$0.26  & 18~14~39.5 & $-$17~52~00 & 35.2 & 14(R, L) & 14(R, L) &  x & 0.1 (R)  \\
IRAS\,18265$-$1517 & 16.87$-$2.16  & 18~29~24.5 & $-$15~15~15 & 16.0 & 1(R) & x  &  x & x  \\
IRAS\,01961$-$0023 & 19.61$-$0.23 & 18~27~38.0 & $-$11~56~36 & 42.0 & x & x &  x & x  \\
IRAS\,02008$-$0013 & 20.08$-$0.13 & 18~28~10.0 & $-$11~28~50 & 46.5 & 0.7(L)& x &  x & x  \\
GRS\,2086$+$048 & 20.86+0.48  & 18~27~25.9 & $-$10~30~24 & 50.0 & x & x &  0.6 (R, L) & x  \\
IRAS\,02886$+$0007 & 28.87+0.06  & 18~43~48.1 & $-$03~35~31 & 103.0 & 2.0(L) & x &  x & 0.5 (R, L)  \\
IRAS\,03060$-$0004 & 30.60$-$0.06  & 18~47~19.9 & $-$02~05~57 & 37.5 &  x & x &  0.4 (L) & x  \\
IRAS\,03124$-$0011 & 31.24$-$0.11  & 18~48~12.4 & $-$01~25~48 & 21.2 & x & x &  x & x  \\
IRAS\,03129$+$0007 & 31.28+0.06  & 18~48~45.2 & $-$01~33~12 & 107.5 & x & x &  x & 0.3 (R, L) \\
IRAS\,03140$-$0026 & 31.40$-$0.26 & 18~49~33.0 & $-$01~29~04 & 85.0 &  x & x &  x & x  \\
Serpens & 31.58+5.38  & 18~29~49.7 & $+$01~15~20 & 9.5 & x & 0.6(R) &  x & x \\
IRAS\,03274$-$0008 & 32.74$-$0.07 & 18~51~21.5 & $-$00~12~11 &  32.3 & x & x &  x & x  \\
%\tablebreak entre las lineas donde se requiere romper la tabla para que quepa en mas paginas 
IRAS\,03313$-$0009 & 33.13$-$0.09 & 18~52~07.3 & $+$00~08~06 & 78.5 &x & x &  x & x  \\
IRAS\,18507$+$0110 & 34.26+0.15  & 18~53~18.7 & $+$01~14~59 & 58.0 & 28(R, L) & 81(R, L) &  x & x  \\
W48 & 35.20$-$1.73  & 19~01~45.5 & +01~13~32 & 32.1 &  4.8(R, L) & 1.9(L) &  x & 1.0  (R, L) \\
IRAS\,03558$-$0003 & 35.58$-$0.03  & 18~56~22.5 & +02~20~27 & 48.9 & 9.5(R, L) & x &  x & x  \\
W51 & 49.49$-$0.39  & 19~23~43.9 & +14~30~27 & 57.0 & 67(R, L) &  3(R, L) &  x & 81.2 (R, L) \\
IRAS\,20126$+$4104 & 78.12+3.63  & 20~14~26.0 & +41~13~32 & $-$4.0 &  0.3(R) & x &  x & x  \\
AFGL\,2591 & 78.89+0.71  & 20~29~24.9 & +40~11~21 & $-$8.8 & x & x &  x & x \\
W70 & 80.87+0.42  & 20~36~52.6 & +41~36~33 & $-$8.0 &  0.6(R, L) & 4.7(R) &  x & x  \\
W75~S & 81.72+0.57  &  20~39~00.9 & +42~22~48 & 1.0 & 13.5(R, L) & 2.3(R, L) &  x & 0.35  (R, L) \\
W75~N & 81.87+0.78  & 20~38~36.8 & +42~37~59 & 7.0 &  130(R, L) &  16(R, L) &  x & 6.4 (R, L) \\
GL2789 & 94.60$-$1.81  & 21~40~00.8 & +50~13~39 & $-$41.0 & 0.8(L) & x &  x & x  \\
IRAS\,21306$+$5539 & 97.52+3.18  & 21~32~11.3 & +55~53~30 & $-$66.5 & x & x &  x & x  \\
IRAS\,21413$+$5442 & 98.04+1.45  & 21~43~00.9 & +54~56~19 & $-$61.0 & 9.5(R) & x &  0.2 (R, L) & x  \\
IRAS\,22176$+$6303 & 106.80+5.31  & 22~19~18.4 & +63~18~45 & $-$8.2 & 2.0(L) & x &  x & x  \\
Cepheus A & 109.87+2.11  & 22~56~18.0 & +62~01~50 & $-$13.8 & 19.0(R, L) &  0.6(R, L) &  x & x  \\
NGC\,7538 & 111.54+0.78  & 23~13~45.3 & +61~28~10 & $-$57.4 &  6.0(R, L) & 1.4(R) &  x & 22.6 (R, L)  \\
IRAS\,00338$+$6312 & 121.30+0.66 & 00~36~47.5 & +63~29~02 & $-$22.8 & x & x & x & x  \\
W3(OH) & 133.90+1.10  & 02~27~04.1 & +61~52~22 & $-$44.3 & 140 (R, L) & 12 (R, L) & 51.1 (R, L) & 19.2 (R, L)  \\
IRAS\,05137$+$3919 &168.06+0.82  & 05~17~13.3 & $+$39~22~14 & $-$21.0 &  0.4(R) & x &  x & x  \\
IRAS\,05168$+$3634 & 170.66$-$0.27  & 05~20~16.1 & $+$36~37~21 & $-$16.0 & 0.3(R) & 1.0(R) &  x & x  \\
S269 & 196.45$-$1.68  & 06~14~37.3 & $+$13~49~36 & 16.0 & 5.5(L) & 2.5(R) &  x & x  \\
\enddata
%% Text for table notes should follow after the \enddata but before
%% the \end{deluxetable}. Make sure there is at least one \tablenotemark
%% in the table for each \tablenotetext.
%\tablecomments{References...}
\tablenotetext{\dag}{Only the flux of the highest peak. 
R stands for right circular polarization and L stands for left circular polarization; 
x means no detection.}
%\tablenotetext{b}{Another sample footnote for table~\ref{tbl-1}}
\end{deluxetable}

\clearpage

%%%%%%%%  Contingency Table  %%%%%%
\begin{deluxetable}{ccccc|c}
\tablecolumns{6}
%\tabletypesize{\scriptsize}
%\rotate
\tablecaption{Contingency Table\label{tab:contingency}}
\tablewidth{0pt}
\tablehead{
& \colhead{1665 MHz} & \colhead{1667 MHz} & \colhead{1612 MHz} & \colhead{1720 MHz} &  \colhead{Total}  
}
\startdata
Detection    & 27 & 16 & 5 & 10 & 58 \\
Non-detection & 14 & 25 & 36 & 31 & 106 \\
\hline
Total        & 41 & 41 & 41 & 41 & 164 \\
\enddata
\end{deluxetable}

\clearpage

\begin{center}
\begin{deluxetable}{crrrrrrrrrrr}
\tabletypesize{\scriptsize}
\tablecolumns{12}
\tablewidth{0pc}
\tablecaption{Pairwise Logistic Regression Models}
\tablehead{
\colhead{}    &  \multicolumn{2}{c}{1665 MHz} &  \colhead{}  & 
\multicolumn{2}{c}{1667 MHz} & \colhead{} & \multicolumn{2}{c}{1612 MHz} & \colhead{}
& \multicolumn{2}{c}{1720 MHz} \\
\cline{2-3} \cline{5-6} \cline{8-9} \cline{11-12}  \\
\colhead{Coefficient} & \colhead{Estimate}  & \colhead{$p$} & 
\colhead{}  & \colhead{Estimate}  & \colhead{$p$} & \colhead{} & \colhead{Estimate}  & \colhead{$p$} & 
\colhead{}  & \colhead{Estimate}  & \colhead{$p$} }
\startdata
  $\beta_{\mbox{\tiny 1665 MHz}}$ & \nodata & \nodata && $2.79 \pm 2.17$ & 0.0118$\dagger$ && $-1.22 \pm  1.92$ & 0.2117 && $1.87 \pm 2.19$ & 0.0932$^*$\\ 
$\beta_{\mbox{\tiny 1667 MHz}}$ & $2.79 \pm 2.17$ & 0.0118$\dagger$ && \nodata & \nodata  && $-1.05 \pm  2.29$ & 0.3690 && $1.74 \pm 1.56 $ & 0.0286$\dagger$ \\ 
$\beta_{\mbox{\tiny 1612 MHz}}$ &  $-1.22 \pm  1.92$ & 0.2117 && $-1.05 \pm  2.29$ & 0.3690 && \nodata & \nodata && $-0.29 \pm 2.32 $ & 0.8078 \\ 
  $\beta_{\mbox{\tiny 1720 MHz}}$ & $1.87 \pm 2.19$ & 0.0932$^*$ && $1.74 \pm 1.56 $ & 0.0286$\dagger$ && $-0.29 \pm 2.32 $ & 0.8078 &&  \nodata & \nodata
\enddata
\tablecomments{Confidence intervals for generalized linear regression model coefficients represented for 95\% confidence calculated from the standard error.}
\tablenotetext{\dagger}{Indicates significance at the $\alpha = 0.05$ level.}
\tablenotetext{*}{ Indicates significance at the $\alpha = 0.1$ level.}
\label{tab:pair_log}
\end{deluxetable}
\end{center}

\clearpage

%%%%%%  Table Global Reg %%%%%%
\begin{center}
\begin{deluxetable}{crrrrrrrrrrr}
\tabletypesize{\scriptsize}
\tablecolumns{12}
\tablewidth{0pc}
\tablecaption{Global Regression Models}
\tablehead{
\colhead{}    &  \multicolumn{2}{c}{1665 MHz} &  \colhead{}  & 
\multicolumn{2}{c}{1667 MHz} & \colhead{} & \multicolumn{2}{c}{1612 MHz} & \colhead{}
& \multicolumn{2}{c}{1720 MHz} \\
\cline{2-3} \cline{5-6} \cline{8-9} \cline{11-12}  \\
\colhead{Coefficient} & \colhead{Estimate}  & \colhead{$p$} & 
\colhead{}  & \colhead{Estimate}  & \colhead{$p$} & \colhead{} & \colhead{Estimate}  & \colhead{$p$} & 
\colhead{}  & \colhead{Estimate}  & \colhead{$p$} }
\startdata
$\beta_{\mbox{\tiny 1665 MHz}}$ & - &- && $2.48 \pm 2.21$  & 0.0283$\dagger$ && $-1.04 \pm 2.23$ &  0.3617 && $1.26 \pm 2.40$ & 0.3031 \\ 
$\beta_{\mbox{\tiny 1667 MHz}}$ & $2.45 \pm 1.89$ & 0.0308$\dagger$ && - & -  && $-0.63 \pm  2.70$ & 0.6476 && $1.33 \pm 1.70$ & 0.1275 \\ 
$\beta_{\mbox{\tiny 1612 MHz}}$ & $-0.94 \pm  2.18$ & 0.4199 && $-0.76 \pm 2.80$ & 0.5940 && - & - && $0.33 \pm 2.63$ & 0.8034 \\ 
$\beta_{\mbox{\tiny 1720 MHz}}$ & $1.15 \pm 3.07$ & 0.3416 && $1.35 \pm 1.86$ & 0.1210 && $0.35 \pm 2.59$ & 0.7935 &&  - & - 
\enddata
\tablecomments{Confidence intervals for generalized linear regression model coefficients represented for 95\% confidence calculated from the standard error.}
\tablenotetext{\dagger}{ Indicates $p$-values significant at the $\alpha = 0.05$ level.}
\label{tab:Log}
\end{deluxetable}
\end{center}

\clearpage

%\begin{comment}
\section{Figures}
%%%%%%%  Figures    %%%%

%\end{comment}

%%%%%%%%%%    End of document  %%%%%%%%%%
\end{document}